\documentclass[preprint,showpacs,preprintnumbers,amsmath,amssymb]{revtex4}

\usepackage{graphicx}
\usepackage{dcolumn}
\usepackage{bm}
\usepackage{amssymb}

\newcommand{\sigone}{$\sigma_{1}(\omega)$}

\newcommand{\x}{$x$}

\newcommand{\w}{${\omega}$}
\newcommand{\bra}[1]{\langle #1|}
\newcommand{\ket}[1]{|#1\rangle}
\newcommand{\braket}[1]{\langle #1\rangle}
\newcommand{\s}{\hspace{.05cm}}

\begin{document}

\title{The Torsional Pendulum as an Example of Electrical anharmonicity}

\author{Jason N. \surname{Hancock}}
\email{jason@physics.ucsc.edu}
\author{Trieau T. \surname{Mai}}
\author{Zack \surname{Schlesinger}}
\affiliation{Physics Department, University of California Santa Cruz, Santa Cruz, CA 95064, USA}

\date{\today}
\begin{abstract}
We demonstrate that an effect other than anharmonicity can severely distort the spectroscopic signatures of quantum mechanical systems. This is done through an analytic calculation of the spectroscopic response of a simple system, a charged torsional pendulum. One may look for these effects in the optical data of real systems when for example a significant rocking component of rigid polyhedra plays a significant role in the lattice dynamics.
\end{abstract}
\maketitle

In a typical spectroscopic experiment, electromagnetic radiation couples to the charge degrees of freedom of a system whose underlying dynamical behavior is described by a Hamiltonian.
In many cases, the behavior of a system near a stable equilibrium can be adequately described by a simple harmonic Hamiltonian, a choice that is often well justified as an expansion in some suitably chosen coordinate about a minimum of a more complicated potential. The potential expansion coordinates are often chosen to be one or more of the Cartesian coordinates $x$, $y$, and $z$, such as is done in the quantum theory of lattice vibrations\cite{ashcroft}.

The coupling of radiation to matter also involves matrix elements of the Cartesian variables and as a result, selection rules arise which forbid optical transitions between vibrational levels which are not adjacent in energy. The effect on spectroscopy is to produce a single peak in the system response at the oscillator's fundamental frequency. The symmetry forbidding the transitions to higher energy levels is imposed by the initial choice of Cartesian expansion coordinate, a choice that in certain situations could be improved upon toward describing the system dynamics. An example of the latter is a system with inherent curvilinear geometrical constraints.

In this manuscript, we show that for a simple and familiar model system where curvilinear motion is inherent, profound effects on the spectroscopic response functions are realized. We exemplify this principle by studying the dynamical response of a torsional pendulum with charge degrees of freedom.

\begin{figure}
\begin{center}
\includegraphics[width=5in]{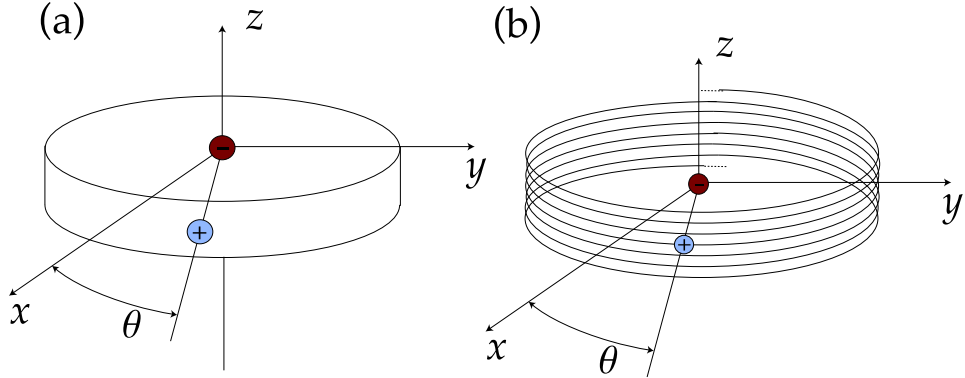}
\caption{(a) A torsional pendulum. (b) A realization of an analogous quantum mechanical system.}
\label{fig:torpic}
\end{center}
\end{figure}


The quantum mechanical problem which is analogous to the classical torsional pendulum is shown in Figure \ref{fig:torpic}b. This situation corresponds to a situation where 360$^\circ$ rotations of the particle wavefunction produces a distinct state, which cannot interfere with the unrotated state. The ``particle on a pig's tail" system shown in Figure \ref{fig:torpic} is can be viewed as a 1D harmonic oscillator wrapped many times around a cylinder. We proceed to investigate the spectroscopic response functions of this harmonic system.

***Quantum Lorentz Oscillator***
To begin we show that, in the absence of damping, a charged linear harmonic oscillator gives rise to response only at the oscillator resonant frequency and hence will have a single peak in the spectrum at this frequency. The Hamiltonian for the linear oscillator is
\begin{equation*}
H=\frac{p^2}{2m}+\frac{1}{2}m \omega_0^2 x^2
\end{equation*}
where $m$ is the particle mass and $\omega_0$ is the resonant frequency of the pendulum. The momentum operator $p$ is conjugate to the position coordinate \x, which can have any value on the \x\ axis. We exploit extensively Dirac's factorization procedure in order to find the current-current correlation function, and hence the linear response of this model system.
The Hamiltonian in quantized form is:
\[H=(a^\dag a+\frac{1}{2})\hbar\omega_0\]
and
\[H\ket{n}=\hbar\omega_0(n+\frac{1}{2})\ket{n}\]
Where $\ket{n}$ is the $n$th harmonic oscillator state and $a^\dag$ and $a$ are the raising and lowering operators of Dirac's theory, which are related to the observables \x\ and $p$ by
\begin{eqnarray*}
x&=&\sqrt{\frac{\hbar}{2m\omega_0}}(a+a^\dag)\\
p&=&-i\sqrt{\frac{m\omega_0\hbar}{2}}(a-a^\dag)
\end{eqnarray*}
The spectrum of this system is a ladder of states in energy separated by $\hbar\omega_0$.

The vector components of the polarization operator are
\[ P_x  =  Q x;\s\s P_y  =  0\]
and the associated current operator can be determined from the Heisenberg equation of motion,
\begin{eqnarray*}
j_x & = & \frac{i}{\hbar}[H,P_x]= \frac{Q}{m}p.
\end{eqnarray*}

The matrix elements we will be interested in are
\begin{eqnarray*}
\label{ }
\bra{n}j_x\ket{m}&=&\frac{Q }{m} (-i\sqrt{\frac{m\omega_0\hbar}{2}}) \bra{n}(a-a^\dag)\ket{m}
\\
&=&\frac{Q }{m} (-i\sqrt{\frac{m\omega_0\hbar}{2}})(\delta_{n,m-1}-\delta_{n,m+1})
\end{eqnarray*}

The optical conductivity, \sigone, which represents the amount of current that is generated by an oscillatory field of frequency \w, is generally related to these matrix elements through
\begin{equation*}
\label{ }
\sigma_1(\omega)=\frac{\pi}{V\omega}\sum_{n\neq0}|\bra{n}\textbf{j}\ket{0}|^2(\delta(\hbar\omega-E_n+E_0)+\delta(\hbar\omega+E_n-E_0)),
\end{equation*}
which for the case of the linear harmonic oscillator, is 
\begin{equation*}
\label{ }
\sigma_1(\omega)=\frac{\pi}{V}\frac{Q^2}{2m}\sum_{n\neq0}(\delta(\omega-\omega_0)+\delta(\omega+\omega_0)).
\end{equation*}
The optical conductivity has a single peak corresponding to absorption of one quantum of radiation of frequency $\omega_0$. When damping is included, this expression generalizes to the Lorentzian line shape, which is encountered in the context of the power absorption in the damped driven harmonic oscillator\cite{marion} and is commonly used in the analysis of optical data where a sharp resonance characterizes the absorption.

The absorption of radiation at only a single frequency is a bit surprising since the harmonic oscillator has energy levels for $n=0,\s1,\s2,\s3,...$, and so one may expect that there be corresponding optical transitions with frequencies $\omega=(E_n-E_0)/\hbar=\omega_0,\s2\omega_0,\s3\omega_0,...$. We shall see below that these transitions are forbidden for the linear harmonic oscillator but need not be in general.

We now repeat this treatment for the torsional oscillator, shown in Figure \ref{fig:torpic}b. One can view this as a 1D harmonic oscillator wrapped on the surface of a cylinder, with the position variable \x\ closely related to the angular coordinate $\theta$. The Hamiltonian for the torsional oscillator is
\begin{equation*}
H=\frac{L^2}{2I}+\frac{1}{2}I \omega_0^2 \theta^2
\end{equation*}
where $I$ is the moment of inertia and $\omega_0$ is the resonant frequency of the pendulum. The angular momentum operator $L$ is conjugate to the (unbounded) angular coordinate $\theta$, and the Hamiltonian describes a harmonic oscillator. We can again exploit extensively Dirac's factorization procedure as before with the appropriate substitutions $x\rightarrow\theta$ and $p\rightarrow L$.

Repeating the procedure outlined above for the linear harmonic oscillator, the vector components of the polarization operator are
\[ P_x  =  Q r \cos\theta\]
\[ P_y  =  Q r \sin\theta\]
and the associated current operators can be determined from the Heisenberg equation of motion,
\begin{eqnarray*}
j_x & = & \frac{i}{\hbar}[H,P_x] =  -\frac{ Q r}{2 I}(L \sin\theta+\sin\theta L)\\
j_y & = & \frac{i}{\hbar}[H,P_y] =  \frac{ Q r}{2 I}(L \cos\theta+\cos\theta L).
\end{eqnarray*}

In the torsional oscillator, we are concerned with taking matrix elements of a nontrivial function of the angle $\theta$, which appears quadratically in the Hamiltonian. This nonlinear dependence of the matrix element on the harmonic degree of freedom will permit transitions between states not adjacent in energy, an effect which is studied widely and has been termed electrical anharmonicity. The main effect of this for spectroscopy is that the conductivity will be a series of peaks at integral multiples of the resonant frequency. We will demonstrate this effect below by calculating the spectroscopic matrix elements for this system.

The matrix elements we will be interested in are of the form
\begin{equation*}
\label{ }
\bra{n}j_x\ket{m}=-\frac{Q r}{2 I }  \bra{n} (L \sin\theta+\sin\theta L)\ket{m}.
\end{equation*}

Cribbing the result from the harmonic oscillator theory,
\begin{eqnarray*}
\label{ }
L=-i\sqrt{\frac{I\omega_0\hbar}{2}}(a-a^\dag),
\end{eqnarray*}
we can write the current matrix elements in terms of matrix elements of the trigonometric functions of $\theta$:
\begin{eqnarray*}
\bra{n}j_x\ket{m} & = & -\frac{Q r}{2 I } (-i\sqrt{\frac{I\omega_0\hbar}{2}}) \bra{n} ((a-a^\dag) \sin\theta+\sin\theta(a-a^\dag))\ket{m} \\
 & = &   i\frac{Q r}{2 I } \sqrt{\frac{I\omega_0\hbar}{2}} (\bra{n+1}\sqrt{n+1}\sin\theta\ket{m}-\bra{n-1}\sqrt{n}\sin\theta\ket{m}
 \\
 & & \hspace{2in}+\bra{n}\sin\theta\sqrt{m}\ket{m-1})-\bra{n}\sin\theta\sqrt{m+1}\ket{m+1}\\
 & = &   i\frac{Q r}{2 I } \sqrt{\frac{I\omega_0\hbar}{2}} (\mathbb{S}_{n+1,m}\sqrt{n+1}-\mathbb{S}_{n-1,m}\sqrt{n}
  \\
 & & \hspace{2in}+\mathbb{S}_{n,m-1}\sqrt{m}-\mathbb{S}_{n,m+1} \sqrt{m+1})
\end{eqnarray*}
where
\begin{equation*}
\label{ }
\mathbb{S}_{n,m}=\bra{n}\sin\theta\ket{m}
\end{equation*}

A nearly identical calculation for the current component $j_y$ gives
\begin{eqnarray*}
\bra{n}j_y\ket{m} & = &  -i\frac{Q r}{2 I } \sqrt{\frac{I\omega_0\hbar}{2}} (\mathbb{C}_{n+1,m}\sqrt{n+1}-\mathbb{C}_{n-1,m}\sqrt{n}
 \\
 & & \hspace{2in}+\mathbb{C}_{n,m-1} \sqrt{m}-\mathbb{C}_{n,m+1}\sqrt{m+1})
\end{eqnarray*}
\begin{equation*}
\label{ }
\mathbb{C}_{n,m}=\bra{n}\cos\theta\ket{m}
\end{equation*}

To take a general approach, we will calculate the matrix elements of the operator
\begin{equation*}
\mathbb{E}_{n,m}=\bra{n}e^{i \theta}\ket{m}
\end{equation*}
and use them to determine those of the trigonometric functions through deMoivre's identity.

Again using a result from the harmonic oscillator theory, and introducing the factor $c$ for notational simplification,
\begin{equation*}
\label{ }
\theta=\sqrt{\frac{\hbar}{2I\omega_0}}(a+a^\dag)=c(a+a^\dag),
\end{equation*}
\begin{equation*}
\label{ }
\mathbb{E}_{n,m}(c)=\bra{n}e^{i c(a+a^\dag)}\ket{m}.
\end{equation*}
The trigonometric matrix elements are then
\begin{eqnarray*}
\mathbb{S}_{n,m}(c) & = & \frac{1}{2 i}(\mathbb{E}_{n,m}(c)-\mathbb{E}_{n,m}(-c)) \\
\mathbb{C}_{n,m}(c) & = & \frac{1}{2}(\mathbb{E}_{n,m}(c)+\mathbb{E}_{n,m}(-c)) 
\end{eqnarray*}

To simplify the calculation of $\mathbb{E}_{n,m}(c)$, one can invoke the Baker-Campbell-Haussdorf theorem\cite{harter}:
\begin{equation*}
\label{ }
e^{A+B}=e^A e^B e^{-[A,B]/2}
\end{equation*}
which holds provided that $A$ and $B$ both commute with their mutual commutator. This is true for $a$ and $a^\dag$, so 
\begin{equation*}
\label{ }
\mathbb{E}_{n,m}=\bra{n}e^{i c a}e^{i c a^\dag}e^{c^2 [a,a^\dag]/2}\ket{m}=\bra{n}e^{i c a}e^{i c a^\dag}e^{ c^2 /2}\ket{m}.
\end{equation*}

We can now expand the exponentials and operate on the bra and ket multiple times with the $a$ and $a^\dag$. Using the identities
\begin{equation*}
\label{ }
\ket{m}=\frac{(a^\dag)^m}{\sqrt{m!}}\ket{0}
\end{equation*}
and
\begin{equation*}
\label{ }
\bra{n}=\bra{0}\frac{(a)^n}{\sqrt{n !}}
\end{equation*}

\begin{eqnarray*}
\mathbb{E}_{n,m} & = & e^{ c^2 /2}\bra{n}e^{i c a} \sum_{k=0}^{\infty} \frac{(i c)^k}{k!} (a^\dag)^k \ket{m}
 \\
& = & e^{ c^2 /2}\sum_{k=0}^{\infty} \bra{n}e^{i c a}  \frac{(i c)^k}{k!} (a^\dag)^k \frac{(a^\dag)^m}{\sqrt{m!}} \ket{0}
 \\
& = & e^{ c^2 /2} \sum_{k=0}^{\infty} \bra{n}e^{i c a}\frac{(ic)^k}{k!}\frac{\sqrt{(m+k)!}}{\sqrt{m!}}\frac{(a^\dag)^{m+k}}{\sqrt{(m+k)!}} \ket{0}
\\
& = & e^{ c^2 /2}\sum_{k=0}^{\infty} \bra{n}e^{i c a}  \frac{(i c)^k}{k! }\frac{\sqrt{(m+k)!}}{\sqrt{m!}} \ket{m+k}
\end{eqnarray*}

Similarly, the $e^{ica}$ can be expanded to act on the bra $\bra{n}$,
\begin{eqnarray*}
\mathbb{E}_{n,m} & = & \sum_{k=0}^{\infty}  e^{ c^2 /2}\bra{n}e^{i c a} \frac{(i c)^k}{k! }\frac{\sqrt{(m+k)!}}{\sqrt{m!}} \ket{m+k}
 \\
& = & e^{ c^2 /2} \sum_{k=0}^{\infty} \sum_{l=0}^{\infty} \frac{(i c)^l}{l!} \bra{n} (a)^l \frac{(i c)^k}{k! }\frac{\sqrt{(m+k)!}}{\sqrt{m!}} \ket{m+k}
\\
& = & e^{ c^2 /2} \sum_{k=0}^{\infty}\sum_{l=0}^{\infty} \frac{(i c)^l}{l!} \bra{0}\frac{(a)^n}{\sqrt{n!}} (a)^l \frac{(i c)^k}{k! }\frac{\sqrt{(m+k)!}}{\sqrt{m!}} \ket{m+k}
\\
& = & e^{ c^2 /2} \sum_{k=0}^{\infty}\sum_{l=0}^{\infty} \frac{(i c)^l}{l!} \bra{0}\frac{(a)^{n+l}}{\sqrt{(n+l)!}} \frac{\sqrt{(n+l)!}}{\sqrt{n!}} \frac{(i c)^k}{k! }\frac{\sqrt{(m+k)!}}{\sqrt{m!}} \ket{m+k}
\end{eqnarray*}

Combining factors and using the orthonormality of the harmonic oscillator states,
\begin{eqnarray*}
\mathbb{E}_{n,m}(c) & = & \frac{e^{ c^2 /2}}{\sqrt{n!m!}} \sum_{l,k=0}^{\infty} \frac{(i c)^{k+l}\sqrt{(l+n)!(k+m)!}}{l! k!} \delta_{l+n,k+m}
\\
& = & \frac{e^{ c^2 /2}}{\sqrt{n!m!}} \sum_{k=0}^{\infty} \frac{(i c)^{2k+m-n}(k+m)!}{k!(k+m-n)!}
\\
& = & \frac{e^{ c^2 /2}(i c)^{m-n}}{\sqrt{n!m!}} \sum_{k=0}^{\infty} \frac{(-c^2)^{k}(k+m)!}{k!(k+m-n)!}\\
& = & \frac{e^{ c^2 /2}(i c)^{m-n}}{\sqrt{n!m!}} S(c^2,m,n)
\end{eqnarray*}
The sum $S(c^2,m,n)$ is a real function and can be expressed in terms of hypergeometric and gamma functions\footnote{$S(c^2,m,n)=   _1F_1(1+m,1+m-n,-c^2) \frac{\Gamma(1+m)}{\Gamma(1+m-n)}$ where $_1F_1(x,y,z)$ is the Kummer confluent hypergeometric function.}. This expression can now be used to give the trigonometric matrix elements of $\theta$ and these are
\begin{eqnarray*}
\mathbb{S}_{n,m}(c) & = & \frac{e^{ c^2 /2}}{\sqrt{n!m!}} S(c^2,m,n) \frac{(i c)^{m-n}-(-i c)^{m-n}}{2 i}\\
 & = & \frac{1}{i}\frac{e^{ c^2 /2}}{\sqrt{n!m!}} S(c^2,m,n) \Bigg\{\begin{array}{lll}
 0  &  m-n & \textrm{even}\\ (ic)^{m-n}  &  m-n & \textrm{odd} 
\end{array}
\end{eqnarray*}
and
\begin{eqnarray*}
\mathbb{C}_{n,m}(c) & = & \frac{e^{ c^2 /2}}{\sqrt{n!m!}} S(c^2,m,n) \frac{(i c)^{m-n}+(-i c)^{m-n}}{2}\\
 & = & \frac{e^{ c^2 /2}}{\sqrt{n!m!}} S(c^2,m,n) \Bigg\{\begin{array}{lll}
 (ic)^{m-n}  &  m-n & \textrm{even}\\ 0  &  m-n & \textrm{odd} 
 \end{array}
\end{eqnarray*}

We restrict our attention at this point to the case of zero temperature. In that case, we are interested in transitions from the ground state, so that $n=0$:
\begin{eqnarray*}
\bra{0}j_x\ket{m} & = & i\frac{Qr}{2I}\sqrt{\frac{I\omega_0\hbar}{2}}(\mathbb{S}_{1,m}+\mathbb{S}_{0,m-1}\sqrt{m}-\mathbb{S}_{0,m+1}\sqrt{m+1})
\end{eqnarray*}

This is zero when $m$ is odd. For $m$ even,
\begin{eqnarray*}
\bra{0}j_x\ket{m}  & = & i\frac{Qr}{2I}\sqrt{\frac{I\omega_0\hbar}{2}}e^{c^2/2}\sum_{k=0}^{\infty}\frac{(-c^2)^k}{k! i}\Big\{\frac{(k+m)!(ic)^{m-1}}{(k+m-1)!\sqrt{m!}}\\
 &&\hspace{1in}+\frac{(k+m-1)!(ic)^{m-1}\sqrt{m}}{(k+m-1)!\sqrt{(m-1)!}}-\frac{(k+m+1)!(ic)^{m+1}\sqrt{m+1}}{(k+m+1)!\sqrt{(m+1)!}}\Big\}
\\
 & = & i\frac{Qr}{2I}\sqrt{\frac{I\omega_0\hbar}{2}}e^{c^2/2}\sum_{k=0}^{\infty}\frac{(-c^2)^k}{k! i}\Big\{\frac{(k+m)(ic)^{m-1}}{\sqrt{m!}}+\frac{(ic)^{m-1}m}{\sqrt{m!}}-\frac{(ic)^{m+1}}{\sqrt{m!}}\Big\}
\\
 & = & i\frac{Qr}{2I}\sqrt{\frac{I\omega_0\hbar}{2}}e^{c^2/2}\sum_{k=0}^{\infty}\frac{(-c^2)^k}{k! i}\frac{(ic)^{m}}{\sqrt{m!}}\Big\{\frac{(k+2m)}{ic}+\frac{c^2}{ic}\Big\}
\\
 & = & i\frac{Qr}{2I}\sqrt{\frac{I\omega_0\hbar}{2}}e^{c^2/2}\sum_{k=0}^{\infty}\frac{(-c^2)^k}{k! i}\frac{(ic)^{m-1}}{\sqrt{m!}}\Big\{k+2m+c^2\Big\} 
 \end{eqnarray*}

We can now sum the series, which gives exponentials:
\begin{eqnarray*}
\bra{0}j_x\ket{m}  & = & \frac{Qr}{2I}\sqrt{\frac{I\omega_0\hbar}{2}}e^{c^2/2}\frac{(ic)^{m-1}}{\sqrt{m!}}\sum_{k=0}^{\infty}\frac{(-c^2)^k}{k! }\Big\{k+2m+c^2\Big\}
\\
& = & \frac{Qr}{2I}\sqrt{\frac{I\omega_0\hbar}{2}}e^{c^2/2}\frac{(ic)^{m-1}}{\sqrt{m!}}\sum_{k=0}^{\infty}\Big\{\frac{(-c^2)^k}{k! }k+\frac{(-c^2)^k}{k! }(2m+c^2)\Big\}
\\
& = & \frac{Qr}{2I}\sqrt{\frac{I\omega_0\hbar}{2}}e^{c^2/2}\frac{(ic)^{m-1}}{\sqrt{m!}}\Big\{-c^2e^{-c^2}+e^{-c^2}(2m+c^2)\Big\}
\\
& = & \frac{Qr}{2I}\sqrt{\frac{I\omega_0\hbar}{2}}e^{-c^2/2}\frac{(ic)^{m-1}}{\sqrt{m!}}2m
\end{eqnarray*}

The analysis of the $\bra{0}j_y\ket{m}$ matrix element follows similarly. The intensities of the transitions are determined from
\begin{eqnarray*}
|\bra{0}j_x\ket{m}|^2 & = &  \frac{Q^2 r^2}{I} \frac{\hbar \omega_0}{2} e^{-c^2} \frac{m^2}{m!} (c^2)^{m-1} \Bigg\{\begin{array}{lll}
1  &  m & \textrm{even}\\ 0  &  m & \textrm{odd} \end{array}\\
|\bra{0}j_y\ket{m}|^2 & = & \frac{Q^2 r^2}{I} \frac{\hbar \omega_0}{2} e^{-c^2} \frac{m^2}{m!} (c^2)^{m-1}\Bigg\{\begin{array}{lll}
0  &  m & \textrm{even}\\  1  &  m & \textrm{odd}  \end{array}
\end{eqnarray*}

\begin{figure}\begin{center}\includegraphics[width=5in]{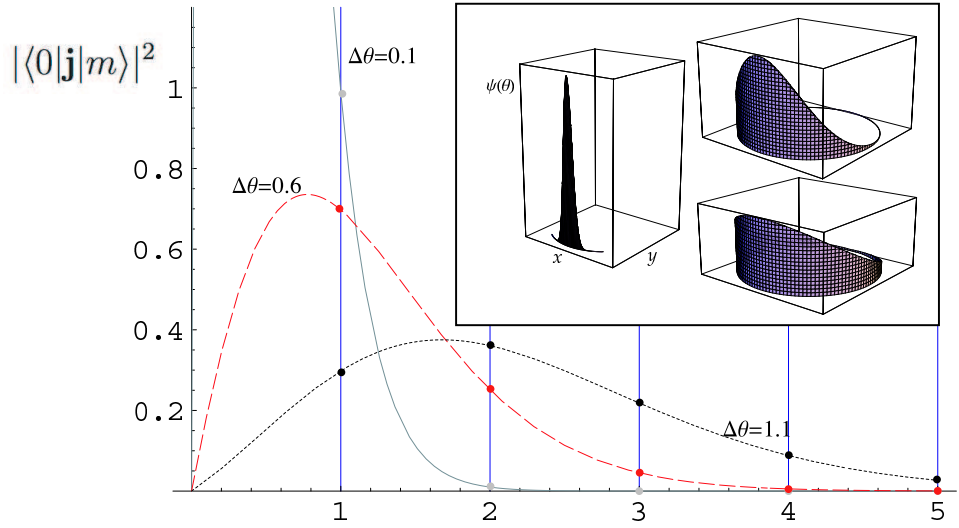}\caption{The matrix elements of the torsional oscillator for three values of angular uncertainty. Only the integral values of $m$ are meaningful. Inset: The ground state wavefunctions corresponding to these three uncertainty values.}\label{fig:tme}\end{center}\end{figure}

Both the relative and absolute intensities are crucially influenced by the parameter $c$, which we have not yet supplied a physical interpretation. To this end, we calculate the uncertainty in angular position of the ground state,
\begin{eqnarray*}
\Delta\theta^2 & = & \braket{\theta^2-\braket{\theta}^2} \\
 & = & c^2\bra{0}(a+a^\dag)^2\ket{0}\\
 & = & c^2\bra{0}(a a^\dag+a^\dag a)\ket{0}\\
 & = & c^2.
\end{eqnarray*}
It seems that the parameter which controls the multiple-peak effect is $\Delta\theta=c=\sqrt{\hbar/2I\omega_0}$, that is the extent to which the wavefunction covers the circle.


We see that the transitions between the ground and excited states are allowed for all excited states in the torsional pendulum, and the distribution of intensities is crucially determined by the angular uncertainty. In the ``stiff pendulum" limit, $\omega_0$ is large and $c\ll1$.
The ground state wave function in this case subtends a small angle and motion along the periphery of the pendulum is well approximated by the appropriate Cartesian coordinate $y=r\sin{\theta}\sim r\theta$.
The $\ket{0}\rightarrow\ket{1}$ transition is by far the strongest, with the other peak exponentially suppressed both as a function of $c$ and $m$. Conversely, in a floppy pendulum, the perpendicular motion is important to the response and also the matrix elements to higher states become appreciable. The crossover between these limits occurs when the angular uncertainty $c$ becomes comparable to 1 radian.

For values of uncertainty $c>$1 radian, the maximum intensity is no longer the $\ket{0}\rightarrow\ket{1}$ transition, but rather the $\ket{0}\rightarrow\ket{2}$ begins to dominate the oscillator strength. Figure \ref{fig:tme} shows this in a plot of $|\braket{0|\textbf{j}|m}|^2$ versus final state quantum number $m$ for several values of $\Delta\theta$.


\begin{figure}
\begin{center}
\includegraphics[width=5in]{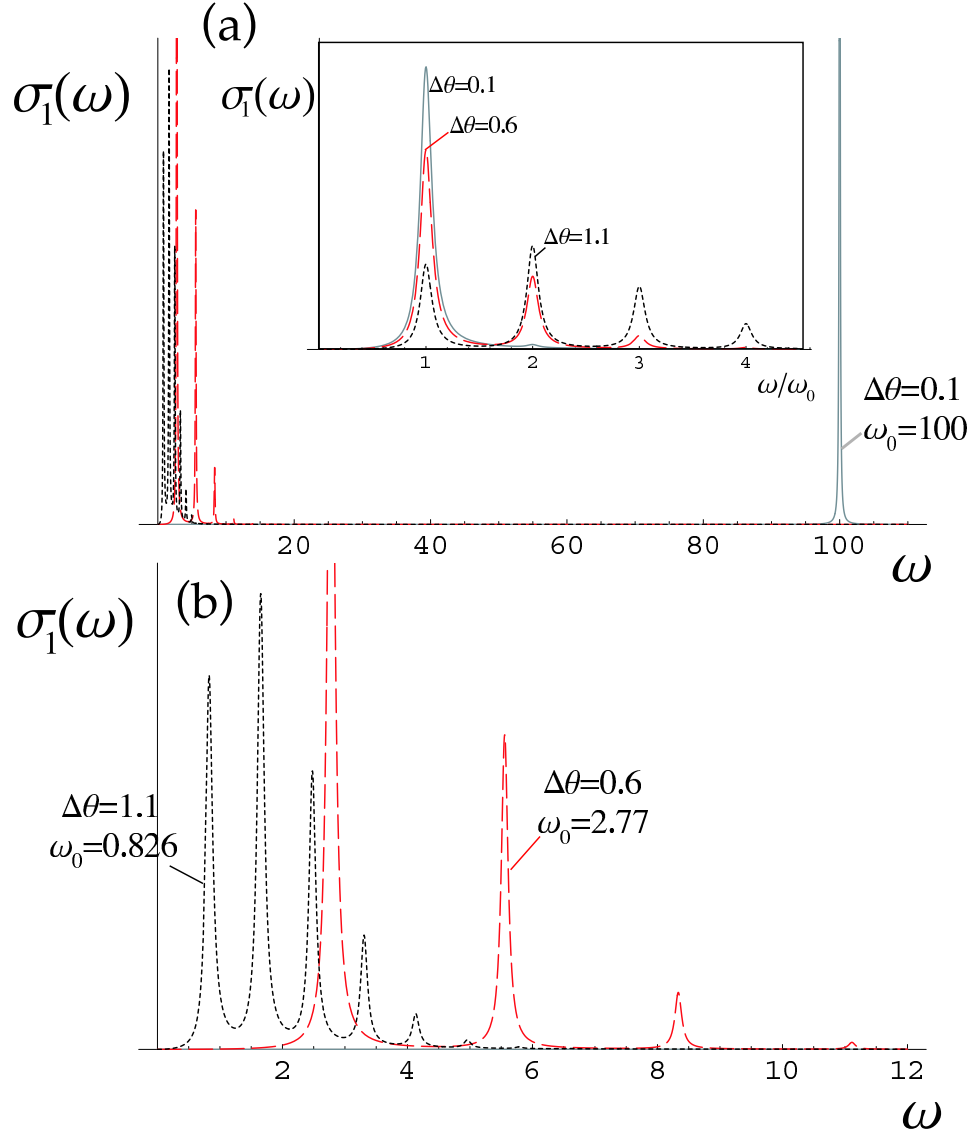}
\caption{The conductivity of the torsional pendulum versus frequency for three values of $c$ (and therefore $\omega_0$). Inset: The same data plotted as a function of final state quantum number.}
\label{fig:condw}
\end{center}
\end{figure}

The diagonal (($\sigma_1^{xx}+\sigma_1^{yy}$)/2) optical conductivity is:
\begin{eqnarray*}
\sigma_1(\omega) &=&\frac{\pi}{V\omega}\sum_{m\neq0}|\bra{m}\textbf{j}\ket{0}|^2(\delta(\hbar\omega-E_m+E_0)+\delta(\hbar\omega+E_m-E_0))\\
& = & \frac{\pi}{V\omega}\frac{Q^2 r^2}{I}\frac{\hbar\omega_0}{2}e^{-c^2}\sum_{m=1}^\infty\frac{m^2}{m!}(c^2)^{m-1}(\delta(\hbar\omega-m\hbar\omega_0)+\delta(\hbar\omega+m\hbar\omega_0))\\
& = & \frac{\pi}{V}\frac{Q^2 r^2}{2I}e^{-c^2}\sum_{m=1}^\infty\frac{m}{m!}(c^2)^{m-1}(\delta(\hbar\omega-m\hbar\omega_0)+\delta(\hbar\omega+m\hbar\omega_0))
\end{eqnarray*}

We can integrate the optical conductivity and find the total oscillator strength:
\begin{eqnarray*}
\int_0^\infty \sigma_1(\omega)  d\omega& = &  \frac{\pi}{V}\frac{Q^2 r^2}{2I}e^{-c^2}\sum_{m=1}^\infty\frac{m}{m!}(c^2)^{m-1}\int_0^\infty \delta(\omega-m\hbar\omega_0)d\omega\\
& = & \frac{\pi}{V}\frac{Q^2 r^2}{2I}e^{-c^2}\sum_{m=1}^\infty\frac{m}{m!}(c^2)^{m-1}\\
& = & \frac{\pi}{V}\frac{Q^2 r^2}{2I}e^{-c^2}\sum_{n=0}^\infty\frac{1}{n!}(c^2)^{n}\\
& = & \frac{\pi}{V}\frac{Q^2 r^2}{2I}=\frac{\omega_P^2}{8}.
\end{eqnarray*}

We can also find $\sigma_2(\omega)$ using the Kramer-Kronig relation\cite{dressel}:
\begin{eqnarray*}
\sigma_2(\omega) &=&-\frac{1}{\pi}\mathcal{P}\int_{-\infty}^\infty\frac{\sigma_1(\omega')}{\omega'-\omega}d\omega'\\
&=&-\frac{1}{\pi}\frac{\omega_P^2}{8}e^{-c^2}\sum_{m=1}^\infty\frac{m}{m!}(c^2)^{m-1}\int_{-\infty}^\infty\frac{\delta(\omega'-m\omega_0)+\delta(\omega'+m\omega_0)}{\omega'-\omega}d\omega'\\
&=&\frac{\omega_P^2}{8}e^{-c^2}\sum_{m=1}^\infty\frac{m}{m!}(c^2)^{m-1}\frac{2\omega/\pi}{\omega^2-(m\omega_0)^2},
\end{eqnarray*}
giving for the complex conductivity ($\sigma=\sigma_1+i\sigma_2$)
\begin{eqnarray*}
\sigma(\omega)&=&\frac{\omega_P^2}{8}e^{-c^2}\sum_{m=1}^\infty\frac{m}{m!}(c^2)^{m-1}(\delta(\omega-m\omega_0)+\delta(\omega+m\omega_0)+i\frac{2\omega/\pi}{\omega^2-(m\omega_0)^2})
\end{eqnarray*}

\begin{figure}
\begin{center}
\includegraphics[width=4in]{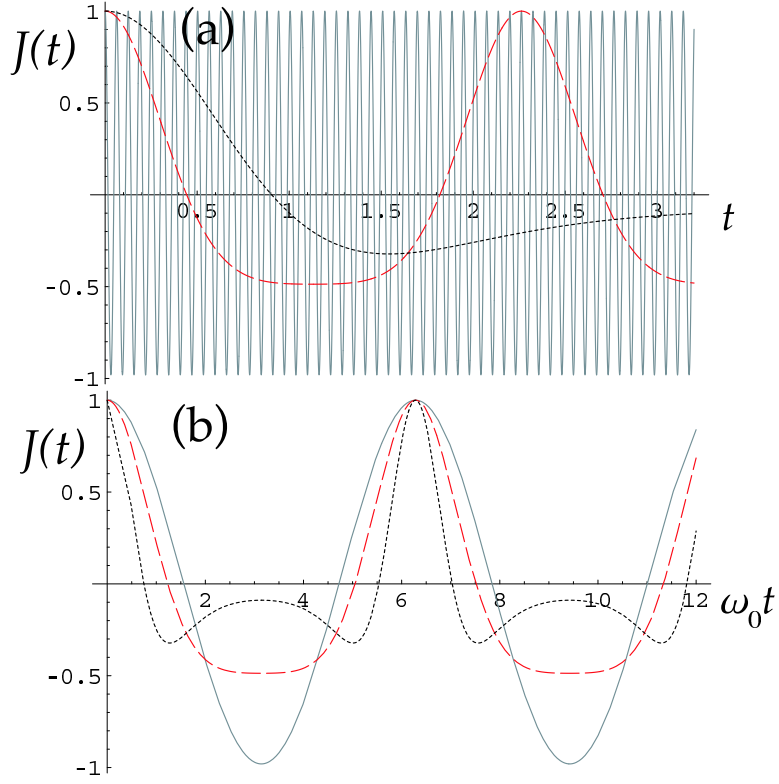}
\caption{(a) The current response $J(t)$ to an impulse pulse at time $t=0$. (b) The same curves as (a), plotted versus the scaled time variable in order to show how the sinusoidal response changes as the pendulum is loosened.}
\label{fig:current}
\end{center}
\end{figure}

Further insight into the physical significance of multiple peaks comes by considering the time-dependent current which arises from the application of a short electric field pulse $E(t)\propto \delta(t)$. In this case, $E(\omega)=E_0=const.$ and for $t>0$,
\begin{eqnarray*}
\label{ }
J(t)&=&\frac{1}{2\pi}\int_{-\infty}^{\infty}e^{-i\omega t}\sigma(\omega)E(\omega)d\omega \\
&=&\frac{E_0}{2\pi}\frac{\omega_P^2}{8}e^{-c^2}\sum_{m=1}^\infty\frac{m}{m!}(c^2)^{m-1}\int_{-\infty}^{\infty}e^{-i\omega t}(\delta(\omega-m\omega_0)+\delta(\omega+m\omega_0)+i\frac{2\omega/\pi}{\omega^2-(m\omega_0)^2})d\omega \\
&=&\frac{E_0}{2\pi}\frac{3\omega_P^2}{4}e^{-c^2}\sum_{m=1}^\infty\frac{m}{m!}(c^2)^{m-1}
\cos m\omega_0 t\\
&=&\frac{E_0}{2\pi}\frac{3\omega_P^2}{4}e^{-c^2}\frac{1}{2}(e^{c^2 e^{-i\omega_0 t}-i\omega_0 t}+e^{c^2 e^{i\omega_0 t}+i\omega_0 t})
\end{eqnarray*}
(for $t<0$, $J(t)=0$, a consequence of causality that is built-in to the Kramers-Kronig relations.)
This current response is shown for three values of $c$ in Figure \ref{fig:current}. For small values of $c$, the response of the system is similar to that of a harmonic oscillator, exhibiting nearly sinusoidal oscillations for $t>0$. Loosening the pendulum (and increasing the angular uncertainty $\Delta\theta$) effects these dynamics considerably.

While the considerations spelled out here are rather idealized, a reasonable place to seek the multiple peak effect in a real system could be the vibrational spectra of nanotubes. When a chiral nanotube is formed from a graphene sheet, phonons propagating along the graphene lattice basis vectors form a twisting pattern around the nanotube axis. The geometry associated with these vibrational degrees of freedom bear similarities to that of Figure \ref{fig:torpic}b. It is feasible that another realization of this effect could be found in solids which support very soft librational phonon modes.

\begin{acknowledgments}
The authors would like to thank B. Sriram Shastry, Trieu Mai, Onuttom Narayan, and Lorenzo Manelli for valuable discussions. Work at UCSC supported by NSF Grant Number DMR-0071949.
\end{acknowledgments}

\section{appendix}
Here we present some calculational details:
\begin{eqnarray*}
j_x & = & \frac{i}{\hbar}[H,P_x]
\\
& = & \frac{i Q r}{2 I \hbar} [L^2, \cos\theta]
\\
& = & \frac{i Q r}{2 I \hbar} (L^2 \cos\theta-\cos\theta L^2)
\\
& = & \frac{i Q r}{2 I \hbar} (L^2 \cos\theta-L\cos\theta L+L\cos\theta L-\cos\theta L^2)
\\
& = & \frac{i Q r}{2 I \hbar} (L [L,\cos\theta]+[L,\cos\theta]L).
\end{eqnarray*}

The commutator in this expression is
\begin{eqnarray*}
[L,\cos\theta] & = & -i \hbar (\frac{\partial}{\partial\theta}\cos\theta-\cos\theta\frac{\partial}{\partial\theta}) \\
 & = & -i \hbar (-\sin\theta+\cos\theta\frac{\partial}{\partial\theta}-\cos\theta\frac{\partial}{\partial\theta}) \\
 & = & i \hbar \sin\theta
\end{eqnarray*}
and so
\begin{eqnarray*}
j_x & = & \frac{i Q r}{2 I \hbar} (L [L,\cos\theta]+[L,\cos\theta]L)
\\
& = & -\frac{ Q r}{2 I}(L \sin\theta+\sin\theta L).
\end{eqnarray*}
A similar expression follows $j_y$

\bibliography{jasonbooks}

\end{document}